\documentclass[aps,prl,superscriptaddress,nobibnotes,nofootinbib,longbibliography]{revtex4-2}

\usepackage{amsmath,amssymb}
\usepackage{graphicx}
\usepackage{tikz}
\usetikzlibrary{calc,arrows.meta,decorations.pathmorphing}
\usepackage{xcolor}

\definecolor{npInk}{HTML}{1F2933}
\definecolor{npMuted}{HTML}{66717C}
\definecolor{npRule}{HTML}{CBD2D9}

\definecolor{npBlue}{HTML}{4C78A8}
\definecolor{npBlueBg}{HTML}{EEF4FA}

\definecolor{npOrange}{HTML}{C9733E}
\definecolor{npOrangeBg}{HTML}{FBF1E9}

\definecolor{npGreen}{HTML}{54846A}
\definecolor{npGreenBg}{HTML}{ECF4EF}

\definecolor{npPurple}{HTML}{7967AA}
\definecolor{npPurpleBg}{HTML}{F1EEF8}

\definecolor{npGrey}{HTML}{66717C}
\definecolor{npGreyBg}{HTML}{F1F3F4}

\definecolor{npRowBlue}{HTML}{F7FAFC}
\definecolor{npRowOrange}{HTML}{FDF9F5}
\definecolor{npRowGrey}{HTML}{F8F9FA}
\definecolor{npRowPurple}{HTML}{FAF8FC}

\usepackage[colorlinks=true,linkcolor=blue,citecolor=blue,urlcolor=blue,filecolor=blue]{hyperref}

\begin{document}

\title{Is higher-order physics different?}
\author{Pablo Villegas}
\email{pablo.villegas@cref.it}
\affiliation{``Enrico Fermi'' Research Center (CREF), Via Panisperna 89A, 00184 -- Rome, Italy}
\affiliation{Instituto Carlos I de F\'isica Te\'orica y Computacional, Universidad de Granada, E-18071 Granada, Spain}

\author{Sandro Meloni}
\email{sandro@ifisc.uib-csic.es}
\affiliation{Institute for Cross-Disciplinary Physics and Complex Systems (IFISC), CSIC-UIB, 07122 -- Palma de Mallorca, Spain}
\affiliation{``Enrico Fermi'' Research Center (CREF), Via Panisperna 89A, 00184 -- Rome, Italy}

\date{\today}

\begin{abstract}
Group interactions are widespread, and higher-order extensions of familiar models display collective phenomena absent from their pairwise baselines, which are routinely offered as evidence of a distinct higher-order physics. We ask whether that claim survives the test that gives \emph{new} a precise meaning in statistical physics: that of universality. Revisiting canonical higher-order models, we argue that what classifies collective behavior are the infrared ingredients that survive at long scales. Arity is not a universality label. Beyond universality, we examine two further questions: whether higher-order structure is a fact about the system or a choice of description, and what data can and cannot tell us about interaction order. Higher-order descriptions remain indispensable when they expose the organizing structure, provide a better mechanistic language, or improve prediction. We close with what should be measured before new phenomena can be claimed, and where higher-order structure already earns its place.
\end{abstract}

\maketitle

\section{A field in expansion}

In less than a decade, higher-order interactions have gone from a technical caveat to a research program, complete with reviews, perspectives and a vocabulary of its own~\cite{battiston2020,battiston2021,lambiotte2019,arruda2024,millan2025,battiston2026}. The motivating observation is unimpeachable: many empirical interactions are polyadic. People meet in groups, neurons fire in assemblies, species compete in ways that depend on who else is present, chemical reactions consume several reagents at once. Representing such systems by hypergraphs or simplicial complexes is natural, and the mathematics that comes with it is beautiful. The problem begins one step later. The observation is about the interactions; the conclusion drawn from it is about the physics---if the couplings are polyadic, the collective behavior must be too. That step is taken often and argued rarely. This Perspective is a critical reading of that program: the question is not whether the models are interesting, but where their novelty lies.


Across the literature, one reads that pairwise networks are \emph{intrinsically limited}, that group interactions generate collective behavior that graphs cannot produce, and that a new physics of higher-order systems is therefore required~\cite{battiston2021,battiston2026}. The evidence usually offered is that simplicial or hypergraph models display a variety of collective behavior: discontinuous transitions, bistability, and hysteresis where their pairwise counterpart displayed a continuous one~\cite{iacopini2019,arruda2023,malizia2025,civilini2024,kim2024}. The observations are real. That they are new does not follow.

The reason is not that the phenomena are unimpressive, but that none of them, taken alone, says what has changed. A discontinuity, a bistable region, a hysteresis loop: none of them had to wait for higher-order interactions~\cite{cai2015,gomezgardenes2011,hansel1993}, and each can be reached by more than one microscopic route---a nonlinearity, a threshold, a memory of past exposures, a second harmonic in a coupling function, or a broken conservation law. Based on current evidence, group interactions are one such route. Whether they are a genuinely new one cannot be read off the outcome; it has to be established case by case, by asking what the group term did to the model, and whether a pairwise rule with comparable ingredients would have done it too. That analysis is seldom carried out. 

A field that produces models faster than it sorts them owes itself a classification. That is the analysis we call for. We begin by re-examining the canonical higher-order models---Ising, contagion, voter, oscillators---asking in each case what the group term actually changed. We then ask whether the distinction is well posed at all: in some cases the same dynamics can be written as a pairwise model with complicated rules, or as a higher-order model with simple ones, and data do not settle which. None of this is an argument against detailed models---they are how one predicts a particular system. We close by asking what higher-order physics have to show, what kind of measurement could show it, and where such a descriptions already earns its place.

\section{What counts as new}

Novelty cannot be asserted without a standard against which it is measured. Statistical physics knows one, and it is worth restating, because it is easy to lose sight of it. Kadanoff put it in a sentence that has not aged: ``All phase transition problems can be divided into a small number of different classes depending upon the dimensionality of the system and the symmetries of the ordered state''~\cite{kadanoff1971}. Wilson's essay for a general audience makes the same point from an other side: the Renormalization Group (RG) works because, at a critical point, the overwhelming majority of microscopic details do not survive coarse-graining~\cite{wilson1979}. The RG is, above all, a machine for forgetting.

A universality class is fixed by a short list: the symmetry of the order parameter, the dimensionality of space, the range of the interactions, the conservation laws, and the presence of disorder. Out of equilibrium the list is equally short.  The Janssen--Grassberger conjecture states that a continuous transition into a single absorbing state, with a one-component order parameter, short-range interactions, no additional symmetry or conservation law and no quenched disorder, belongs to the directed-percolation (DP) class~\cite{janssen1981,grassberger1982, hinrichsen2000,odor2004}. The payoff of such a statement is enormous, and it is worth remembering what it looks like: the onset of sustained turbulence in quasi-one-dimensional Couette flow---a problem with no obvious relation to lattice contagion---displays the critical exponents of $(1+1)$-dimensional DP~\cite{lemoult2016}.
That is the strength of universality.

There is also a historical precedent for the situation higher-order interactions have generated. In the early 1970s the theory of dynamic critical phenomena faced its own explosion of models, each with its own couplings and conservation laws. The field did not answer it with a catalog; it answered with the classification of Hohenberg and Halperin, models A through J, organized by order-parameter symmetry and by what is conserved~\cite{hohenberg1977}. Fifty years later, that table is still how one orients oneself. A proliferation of models is not a problem in itself. It becomes one when nobody asks which of the distinctions among them survive at long scales.

Interaction order can matter at three distinct levels: whether the coupling can be reduced to pairwise terms, when a higher-order representation offers advantages a pairwise one does not, and whether it can be identified from data. Claims of new higher-order physics usually appeal instead to universality, which is precisely the property least often checked.

\section{The canon revisited}
\label{sec:models}

\begin{figure*}[t]
\centering
\includegraphics[width=\textwidth]{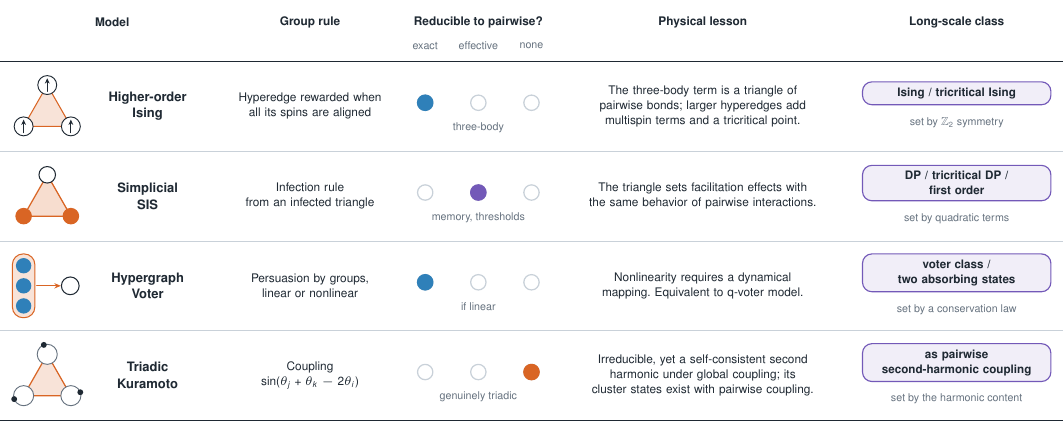}
\caption{\textbf{The canon revisited.}
Canonical models with group interactions: the higher-order Ising model~\cite{robiglio2025}, simplicial SIS contagion~\cite{iacopini2019,meloni2026}, hypergraph voter dynamics~\cite{llabres2026,castellano2009} and triadic Kuramoto coupling~\cite{skardal2019,skardal2020,hansel1993}. The middle column distinguishes exact microscopic reduction to pairwise terms (blue), equivalence only at the level of the effective theory (purple) and microscopic irreducibility (orange). Microscopic reducibility and long-scale class are independent: in every case the infrared verdict is set by the operator content of the effective theory---symmetries, conservation laws, slow fields, multicritical tuning---not by interaction arity. Similar phenomenology does not imply microscopic equivalence, and microscopic irreducibility does not imply a new universality class.}
\label{fig:canon}
\end{figure*}

The higher-order program has, in part, re-run the canon. Take the models on which the statistical physics of collective behavior was built---Ising, contagion, the voter model, phase oscillators---add interactions among three or more units, and see what changes. The exercise has been carried out carefully. What is striking is the recurrence of a small set of macroscopic outcomes. In what follows we take the four models one at a time, and in each case ask what the group term actually changed (see Figure~\ref{fig:canon}).

\emph{\textbf{Ising}.} The higher-order Ising model of Robiglio \emph{et al.} rewards a hyperedge when all its spins are aligned and therefore preserves spin inversion~\cite{robiglio2025}. With three-spin hyperedges the transition is continuous in mean field. Here the ``three-body'' term is in fact exactly reducible: for three Ising spins, $2\delta(s_1{=}s_2{=}s_3)=\tfrac12(1+s_1s_2+s_2s_3+s_3s_1)$, so each hyperedge is, up to a constant, a triangle of pairwise bonds of half the coupling. The Supplemental Material of Ref.~\cite{robiglio2025} makes this reduction explicit and contrasts it, on the same hyperedges, with the conventional $p$-spin rule: the symmetry-preserving effective field is linear in $m$, whereas the odd three-spin product generates an $m^2$ term and breaks spin inversion. Correspondingly, the former transition is continuous and the latter abrupt. Same arity, same substrate; different symmetry and bifurcation. 

Irreducibility returns for larger hyperedges. For four spins the aligned-state interaction contains a genuine $s_1s_2s_3s_4$ term while preserving spin inversion; by lowering the quartic coefficient of the Landau free energy, this can drive the standard tricritical route to a first-order transition. Whether it does depends on the mixture and strength of the interaction orders. For the three-body model, the Georges--Yedidia correction shifts the critical point while retaining the mean-field critical exponent, consistent with the authors' statement that the universality class is unchanged~\cite{robiglio2025}.

\emph{\textbf{Contagion}.} In mean-field simplicial contagion a susceptible node acquires an additional infection channel when the other two nodes of a triangle are simultaneously infected~\cite{iacopini2019}, producing bistability and hysteresis. At mesoscopic scales the dynamics takes the form~\cite{meloni2026}: $\partial_t \rho =-a\rho + b\,\rho^2 - c\,\rho^3 + D\nabla^2\rho + \sqrt{\rho}\,\eta$, where the 2-simplex contributes to the quadratic coefficient $b$, while higher-order simplices generate progressively higher powers of the density. 

A distinct hypergraph contagion model makes the same separation from another direction. Burgio \emph{et al.} show that the three-body infection rate affects the linear stability of the absorbing state only when three- and two-body interactions overlap on the same triangles, and that varying this overlap can shift the invasion threshold and alter the bifurcation structure~\cite{burgio2024}. At the effective level, however, neither a coefficient nor a bifurcation uniquely records its microscopic origin. The same nonlinear structure can arise from facilitation, thresholds or memory of previous exposures, without any simultaneous group contact at all~\cite{dodds2004,bizhani2012}.

The generalized epidemic process, built from local pairwise contacts and an additional weakened state, reaches tricritical dynamic percolation~\cite{janssen2004}; cooperative pairwise contagions can likewise display abrupt or hybrid transitions~\cite{cai2015}. Conversely, introducing permanent immunity changes the absorbing-state structure itself: the recovered field carries memory and takes the problem from directed to dynamic isotropic percolation~\cite{grassberger1983}.

\emph{\textbf{Voter}.} On hypergraphs, the reduction is exact: a broad class of social-impact rules maps onto pairwise dynamics on a weighted projected network, and for linear rules the macroscopic behavior is that of the ordinary voter model~\cite{llabres2026,neuhauser2020}. Nonlinear rules do behave differently, but that case already has a name and a history: the $q$-voter model, in which $q$ neighbors are consulted at once, was analyzed in 2009~\cite{castellano2009}, and on well-connected hypergraphs the nonlinear voter model on the unweighted projection still reproduces the main trends~\cite{llabres2026}. What nonlinearity changes is not the arity but a conservation law. The linear voter model conserves the average magnetization and sits in the voter class~\cite{dornic2001}; nonlinear rules break that conservation, and the voter point becomes a special point of the general theory of two symmetric absorbing states~\cite{alhammal2005}.

\emph{\textbf{Oscillators}.} Triadic phase coupling is the most interesting case because here the microscopic interaction is genuinely irreducible: it cannot be written as a sum of two-body coupling functions. Its structure is nonetheless transparent. The attraction between $i$ and $j$ is modulated by $\cos(\theta_k-\theta_i)$, which, under certain conditions, introduces repulsive effects. For global coupling the sum can be performed exactly, $\frac{1}{N^2}\sum_{j,k}\sin(\theta_j+\theta_k-2\theta_i)=r^2\sin 2(\psi-\theta_i)$, where $re^{i\psi}=N^{-1}\sum_j e^{i\theta_j}$. The irreducible triadic interaction has therefore become a self-consistent second-harmonic locking force. The $\pi$-periodicity of this force gives each locked oscillator two phase branches separated by $\pi$, underpinning the antipodal clustering and extensive multistability of the purely triadic model, together with its continuum of abrupt desynchronization transitions~\cite{skardal2019,zhang2024}. Neither feature is, however, uniquely higher-order. Two-cluster states arise in strictly pairwise oscillator models once a second harmonic is included in the coupling function~\cite{hansel1993}, and discontinuous synchronization itself occurs in pairwise Kuramoto models for suitable frequency distributions or degree--frequency correlations~\cite{pazo2005,gomezgardenes2011}. When pairwise and higher-order couplings are combined, the same nonlinear feedback can also generate hysteretic abrupt synchronization~\cite{skardal2020}. The triad is irreducible; its macroscopic phenomenology is not unique to it.

\begin{figure*}[b]
\centering
\includegraphics[width=\textwidth]{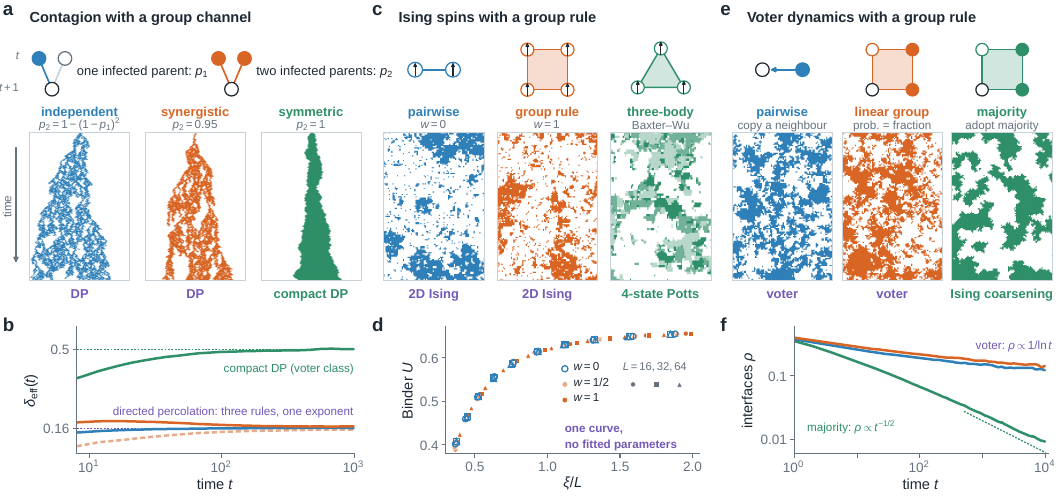}
\caption{\textbf{Group interactions leave the long-scale behavior alone; a new symmetry or conservation law does not.}
(a)~Domany--Kinzel contagion~\cite{domany1984}: a site is infected with probability $p_1$ by one infected parent and $p_2$ by two, the group channel. Critical space--time clusters for independent [$p_2=1-(1-p_1)^2$] and synergistic ($p_2=0.95$) transmission are statistically alike; at the symmetric endpoint $(p_1,p_2)=(1/2,1)$ the active and inactive states become equivalent and the transition is compact DP~\cite{essam1989}.
(b)~Effective survival exponent at criticality: antagonistic ($p_2=0$), independent and synergistic rules all converge to directed percolation (DP), the symmetric point to compact DP; critical points and exponents from series expansions~\cite{jensen1999}.
(c)~Equilibrium configurations at criticality of the square-lattice Ising model with nearest-neighbor bonds only ($w=0$); of the same lattice with a plaquette rewarded only when its four spins agree ($w=1$); and of the Baxter--Wu model~\cite{baxter1973}, whose three-spin interaction has four ground states (white and three shades of green).
(d)~Binder cumulant $U$ versus $\xi/L$ for $L=16,32,64$ and $w=0,1/2,1$: the three models fall on one curve.
(e)~Opinion dynamics on the square lattice from a random initial condition ($t=50$). Copying a random neighbor (pairwise) and adopting an opinion with probability equal to its fraction in a plaquette (linear group rule) give the same rough domains without surface tension~\cite{dornic2001}, as the exact reduction of linear group rules to the projected voter model requires~\cite{llabres2026,neuhauser2020}; adopting the plaquette majority, a nonlinear group rule, breaks the conservation of the mean opinion and yields smooth, curvature-driven domains~\cite{bray1994}, the majority-vote rule lying in the Ising class~\cite{deoliveira1992}.
(f)~Density of interfaces: $\rho\propto 1/\ln t$ for both voter rules~\cite{frachebourg1996,dornic2001}, a power law approaching $t^{-1/2}$ for the majority rule~\cite{bray1994}. Simulations illustrate established results; all critical points and exponents are taken from the literature.}
\label{fig:universality}
\end{figure*}

\emph{\textbf{The lesson was already there}.} Many-body couplings are not a discovery of the last decade. It is worth recalling what the literature already says about them. Three-body dispersion forces between neutral atoms have been known since Axilrod and Teller~\cite{axilrod1943}; they are real, measurable, and can be essential for the thermodynamics of simple fluids, yet they leave the liquid--gas critical point in the Ising class. Plaquette interactions do produce something qualitatively different---a transition with no local order parameter~\cite{wegner1971}---but that follows from the local gauge invariance they respect, not from the fact that it has four legs.

The decisive precedents sit on the triangular lattice. Baxter and Wu solved the Ising model with a purely three-spin interaction and found the exponents not of Ising but of the four-state Potts model~\cite{baxter1973}: the three-spin term gives the ground state a fourfold degeneracy that the two-spin model lacks. It did produce that symmetry; but the class it reaches is the one the four-state Potts model reaches with pairwise couplings. Keep the same product on the same lattice but retain triangles of one orientation only, and the symmetry structure changes completely: the Newman--Moore model possesses symmetries that flip spins on fractal subsets of the lattice and has no finite-temperature transition, only glassy dynamics without quenched disorder~\cite{newman1999}. Same arity, same lattice, same product; a different symmetry structure, and a different physics.

The eight-vertex model provides the complementary case: two Ising copies coupled by a four-spin term, with continuously varying exponents generated by the marginal product of their energy densities~\cite{baxter1971,kadanoff1971b}, and generating the Ashkin--Teller line of fixed points. Fixed lines are no many-body privilege, as the pairwise XY model shows. Disorder repeats the lesson: mean-field $p$-spin glasses with $p\ge3$ freeze discontinuously, unlike the $p=2$ spin glass~\cite{derrida1980,gross1984,kirkpatrick1987}, and so does a pairwise $q$-state Potts glass with $q>4$ from pairwise random couplings~\cite{gross1985}. What survives is the organization of the free-energy landscape, not the number of spins in a microscopic term.

The moral is sharp and general: what matters at long scales is the operator content a many-body term generates---its symmetry, and its relevance at the fixed point---not the number of legs at the vertex.

\emph{\textbf{The pattern}.} Read together, the four cases separate three questions that the literature tends to merge: whether the microscopic rule is reducible, what collective phenomenology it produces, and what survives coarse-graining (see Fig.~\ref{fig:universality}). The symmetric three-spin Ising interaction and the linear voter rule are exactly reducible; simplicial contagion generates familiar nonlinear operators in the mesoscopic theory; triadic phase coupling remains microscopically irreducible.  Similar macroscopic behavior can arise through very different microscopic routes. Discontinuities, bistability, and multistability appear in all of them, and in their pairwise counterparts. They cannot be the signature of arity.

\section{A matter of representation}
The second level at which arity can matter is representation, and there the question is one of fact rather than taste. Graphs are not ``intrinsically limited to pairwise interactions''~\cite{battiston2021}. Factor graphs, bipartite encodings and auxiliary-field decouplings have represented polyadic dependencies for decades, and in this sense hypergraphs are another way of writing graph-based encodings~\cite{peixoto2026}. Where two descriptions are exactly equivalent, the question that remains is not which of them is true but which is the more convenient, and what the convenience costs.

\begin{figure*}[hbtp]
\centering
\includegraphics[width=\textwidth]{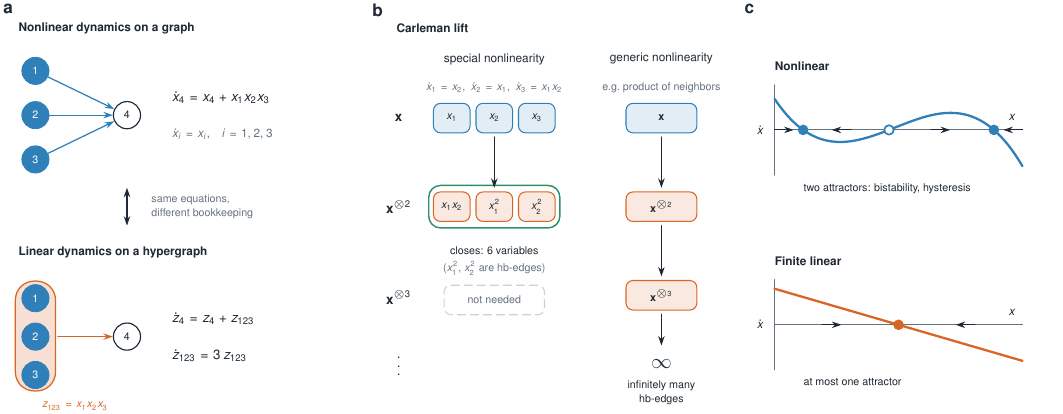}
\caption{\textbf{Representation is not mechanism.}
(a)~A nonlinear graph dynamics and its exact linear lift on a higher-order state space, where the hyperedge variable is $z_{123}=x_1x_2x_3$.
(b)~Carleman lift~\cite{lacasa2026}. For special nonlinearities the lifted space closes after a few steps; monomials with repeated indices ($x_1^2$, $x_2^2$) are states of hyperedges with repeated vertices. For generic nonlinearities the lift never closes, and an exact linear representation requires infinitely many edges.
(c)~A nonlinear flow can support multiple isolated stable attractors and hysteresis; a finite-dimensional autonomous linear system cannot. In an exact lift this complexity is transferred to the enlarged state space.}
\label{fig:representation}
\end{figure*}

Two constructions map the descriptions onto each other. In one direction, the reducibility results rewrite a dynamics on a hypergraph as a nonlinear dynamics on the projected graph: where this has been carried out---linear and nonlinear consensus, social-impact and voter rules---the evolution of each node becomes a nonlinear function of its neighbors~\cite{neuhauser2020,sahasrabuddhe2021,llabres2026}. In the other, Carleman linearization embeds a polynomial dynamics on the vertices of a graph, $\dot x_i = F_i(\mathbf{x})$, into a linear dynamics on a richer structure, at a price we come to below~\cite{lacasa2026}: one lifts the state space, $\mathbf{x}\to(\mathbf{x},\mathbf{x}^{\otimes 2},\dots)$, and monomials such as $x_1x_2x_3$ become the states of hyperedges. Both constructions are recent, and how far each extends is an open program rather than a closed result. What they already establish, in the cases where both apply, is that the two descriptions are not rivals but the same dynamics, with the nonlinearity booked either to the equations or to the structure (Fig.~\ref{fig:representation}a).

The bookkeeping is not free, and the bill is informative. Take the lift from a polynomial dynamics on a graph to a linear one on a hypergraph~\cite{lacasa2026}: it closes after finitely many steps only for special nonlinearities; for generic polynomial couplings, and for analytic nonlinearities more generally, it is infinite-dimensional (Fig.~\ref{fig:representation}b,c). It has to be, since a finite-dimensional autonomous linear system supports no multiple isolated attractors, no coexisting basins and hence no hysteresis, and no chaos. Bistability and multistability do survive the embedding, but they survive in the enlarged state space and its constraints rather than in an evolution law. Linearizing does not remove the complexity. It moves it into the representation, at the price of infinitely many edges.

Which book to keep is, as Lacasa observes, a matter of explanatory commitment~\cite{lacasa2026}. Dynamicists are trained to see coupled equations and no graph at all; network scientists are trained to see structure, and to read ``more is different'' as a statement about topology. Anderson, for what it is worth, meant a statement about broken symmetry. Nor does parsimony settle it: nobody asks whether a chaotic map is really its orbits or really its Perron--Frobenius operator. Representations are chosen for what they let one compute. A linear higher-order representation may well be the right tool for controlling a nonlinear network; it is not, for that reason, a discovery about nature.

A higher-order operator can also resolve a time scale that its pairwise counterpart does not, and that is worth something even when the physics is unchanged. In a study of synergistic spreading of our own, the relaxation scales of a higher-order Laplacian locate the parameters at which the contagion displays an extended, Griffiths-like critical regime~\cite{lucarini2026}. The regime itself is not higher-order: it survives when the group rule is replaced by a pairwise threshold. The operator did not create the phenomenon; it made it visible, and told us where to look.


\section{Can higher-order interactions be inferred?}
The third level is identifiability: every parameter added to a model should be recoverable from data. Bona-fide higher-order interactions thus require seeing them in data, and a growing program sets out to do so. Hypergraphs and simplicial complexes have been reconstructed from contagion and Ising time series~\cite{wang2022}, from the trajectories of coupled dynamical units~\cite{malizia2024}, by sparse regression over libraries of nonlinear terms~\cite{delabays2025,su2026}, and from the stochastic increments of high-dimensional recordings~\cite{tabar2024,santoro2023}. Within their assumptions these methods work: with known coupling functions, pairwise and three-body phase interactions can be told apart in controlled benchmarks~\cite{su2026}. The trouble lies in the assumptions. Data constrain the dynamics; by themselves they do not decide where its nonlinearity should be booked, and a reconstructed hyperedge is a statement about the model class one has chosen to fit. Most methods presuppose that class and few test it against alternatives, even though the question is one of model selection~\cite{lambiotte2019}. Peixoto \emph{et al.} argue that existing reconstructions have not yet accounted for model complexity, and so have not established a hypergraph as the more plausible description of any real system~\cite{peixoto2026}.

\begin{figure*}[t]
\centering
\includegraphics[width=\textwidth]{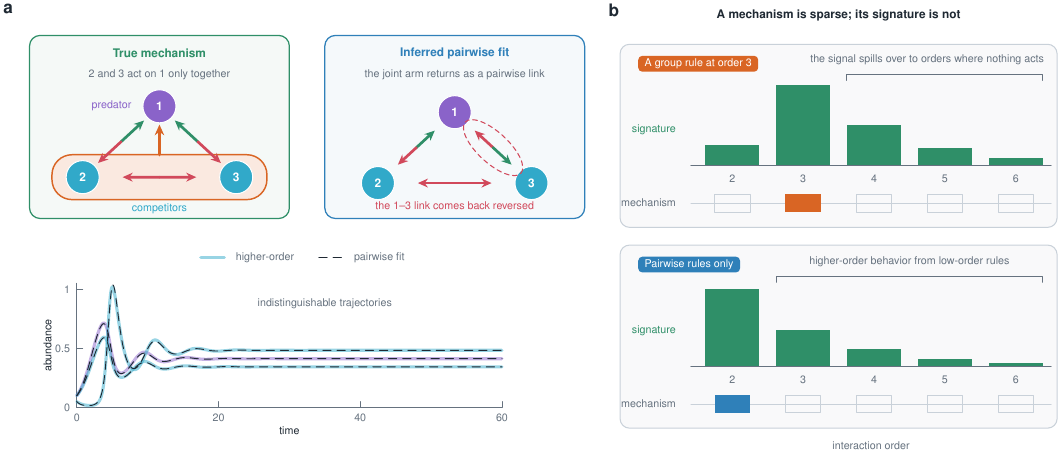}
\caption{\textbf{What data can and cannot see.}
(a)~A three-species community in which species 1 is affected by 2 and 3 only jointly (orange arm); arrowheads carry the sign of each effect, green positive and red negative. The best pairwise fit has no group term, so the joint arm reappears as an ordinary link between 1 and 3, with the sign reversed---a competitor is promoted to predator---while the fitted trajectories stay indistinguishable from the true ones~\cite{callejasolanas2026}.
(b)~A mechanism is sparse; its signature is not. In each panel the green bars give the strength of the statistical signature of dependence at successive interaction orders, and the strip below marks the orders at which a rule actually acts---filled where one does, open where none does. A single group rule at order three leaves a signature that spills over to orders at which nothing acts~\cite{danovski2026}; a system governed exclusively by pairwise rules still displays dependencies among three and more units~\cite{robiglio2025b}. Higher-order \emph{behaviors} and higher-order \emph{mechanisms} are therefore independent~\cite{rosas2022}: synergy answers whether a group mechanism is present, it does not say at which order it acts.}
\label{fig:inference}
\end{figure*}

Ecologists have been here before. Whether higher-order interactions ``really exist'' in competitive communities was asked in 1981~\cite{pomerantz1981,case1981}, and the long debate that followed~\cite{billick1994} has a modern answer: leave out the resources through which species actually compete, and non-additive effects of density appear on their own. They are a property of the phenomenological model, not of the community~\cite{letten2019}. That debate taught ecologists to separate interaction modifications that pairwise experiments already resolve, such as nonlinear density dependence, from those that appear only when three or more competitors meet, and to treat the latter, when a fitted model requires them, as evidence that the system is being described too simply---a question to be settled by independent measurements~\cite{kleinhesselink2022}. The converse has also been tested. Abundance time series generated by specific higher-order Lotka--Volterra dynamics can be reproduced by effective pairwise models: along a trajectory, the higher-order contributions can be absorbed into effective pairwise coefficients~\cite{callejasolanas2026} (Fig.~\ref{fig:inference}a). The fits predict well and explain badly: the inclusion of higher-order effects introduces ecosystem roles that are not there. Epidemics finally tell the same story; pairwise SIS dynamics with time-dependent rates likewise reproduce hypergraph contagion in both transient and stationary regimes~\cite{tan2026}. 

Statistical signatures narrow the distance between the data and the rule, but they do not close it. Higher-order \emph{behaviors}---dependencies among the states of three or more units---must be distinguished from higher-order \emph{mechanisms}, the rules that generate them~\cite{rosas2022}. Neither the distinction nor the difficulty is new. Correlation functions of $n$ points have been the currency of statistical mechanics since the cluster expansions of Ursell and Mayer~\cite{ursell1927,mayer1937}, and systems of pairwise forces have always had them. In a Gaussian theory every $n$-point function is assembled from pairwise contractions; coarse-grain a strictly pairwise Ising model and the effective action acquires a quartic term coupling four fields. Neither is read as evidence of a four-body force. The newer information-theoretic measures---the $O$-information and its relatives---sharpen the old distinction without dissolving it: they resolve, order by order, how much of the dependence within a group of variables is redundant and how much is synergistic~\cite{rosas2019,santoro2023}, and it is with them that the difficulty has been made quantitative rather than merely suspected. Synergy does respond to group mechanisms, but systems with only low-order mechanisms display higher-order behaviors~\cite{robiglio2025b}, and in simplicial contagion the signature spills over to orders at which no mechanism acts~\cite{danovski2026}: a mechanism is sparse in interaction order, its signature is not (see Fig.~\ref{fig:inference}b).

The measures are not useless; they answer a narrower question than the one being asked of them. A positive synergy tells us that the three units are doing something together that none of the three pairs does on its own. It does not tell us that a three-body rule is what makes them do it. Settling that requires a comparison: fit a model with pairwise rules only, and give it the same advantages the higher-order model enjoys---the same memory of past states, the same unobserved inputs, the same freedom in fitting its parameters. If that model produces the same signal, what was measured is emergent behavior and not a higher-order mechanism: the whole does something its parts do not, which is what emergence has always meant.

There is a last issue, and it concerns scale rather than comparison. Inference is performed at a resolution, and the resolution is part of the answer. Decimating a nearest-neighbor Ising model generates four-spin couplings at the first step~\cite{maris1978}, and phase reduction turns pairwise coupling into non-pairwise phase interactions beyond leading order~\cite{leon2019,bick2024,battiston2026}: a hyperedge
inferred at one scale need not be the arity of anything at another. The tools for a conditional answer exist~\cite{peel2022,peixoto2025,musciotto2021,neuhauser2024,lucas2026}; what they ask is that a reconstruction state what it assumed, at what resolution, and against what lower-order control~\cite{callejasolanas2026}. With those declarations, inference is a powerful instrument. Without them, it finds in the data the model one put into the fit.

\section{Two questions, two currencies}

There is an objection to everything we have said so far, and we believe it is a good one. Not all modeling aims at universality. A great deal of excellent science asks a different question: not \emph{which class does this transition belong to}, but \emph{what will this particular system do, and what happens if I intervene on it}. For that second question, the microscopic detail is not noise to be integrated out---it is the answer.

The simplest models exist to expose what is general; detailed models exist to predict what is particular. Both enterprises are legitimate. They are paid in different currencies, and trouble starts only when a debt in one is settled in the other. Group interactions are often indispensable in this second sense. Gene regulation is combinatorial by construction, and thermodynamic models built on multi-factor occupancy states predict expression levels quantitatively~\cite{bintu2005}; interactions that do not decompose into pairs shape protein fitness landscapes and the coexistence of competing species~\cite{poelwijk2019,mayfield2017,grilli2017}. In materials science, three-body potentials of the Stillinger--Weber type are what allow a simulation of silicon to reproduce its tetrahedral structure at all~\cite{stillinger1985}---while three-body forces of the Axilrod--Teller type, as recalled above, move the critical point of a fluid without moving it out of the Ising class. That juxtaposition is our whole argument in miniature.


It is worth adding that within this second enterprise the question ``are higher-order terms needed?'' is empirical, and has a measurable answer.  The necessity of higher-order terms is a model-selection statement about a given system at a given resolution, settled by out-of-sample prediction, and it can come out either way. We would therefore not ask a systems biologist to compute a critical exponent, and we would not accept a critical exponent as evidence about a promoter. What we would ask of both is that they state which question they are answering. Our complaint about the higher-order literature in statistical physics is not that it builds detailed models---detailed models are how one predicts a particular system, and that is a worthy thing to do. It is that detailed models are built and then credited with the reward reserved for the other enterprise: a claim about generality, about what every system in a class must do. That is the debt settled in the wrong coin.

\section{What to measure, what to ask, and where to look}
There is an elephant in the higher-order room, and the three levels discussed here lead to it: reducibility, representation and inference all rest on a comparison with a pairwise description. To our knowledge, that comparison has never yet gone against it.

Our own field suggests where a case of that kind would come from. Universality was measured before it was explained: in 1945 Guggenheim showed that the coexistence curves of eight simple fluids collapse onto a single curve with an exponent near $1/3$ rather than the $1/2$ of van der Waals~\cite{guggenheim1945}, and the classes were built on measurements of that kind long before the renormalization group accounted for them. The same standard still works: the DP class was confirmed not by another lattice model but by measuring the exponents of a turbulent flow~\cite{lemoult2016}.

What is scarce for arity is the Guggenheim evidence: an exponent or a scaling function, measured in a real system, that a pairwise model with the same infrared ingredients cannot fit. We know of no attempt of this kind. The empirical tests we do have---in neural populations, drug combinations, plant communities---answer the predictive question rather than the universality one, and they come out either way~\cite{schneidman2006,katzir2019,mayfield2017}. Group structure itself is not in doubt: high-resolution proximity data resolve social gatherings directly and find them organized around stable cores~\cite{sekara2016}. Still, all is evidence about structure, not about the collective behavior it produces.

Until such evidence exists, we do not know which higher-order variables deserve to be kept. Two questions must be distinguished. The scaling observation just described would identify new collective physics. Evidence for a genuinely higher-order \emph{mechanism} requires, in our opinion, an \emph{intervention} rather than an observation, designed so that no pairwise account with the same inputs could follow it. Neither implies the other, and only their conjunction would be decisive. Data themselves can also be coarse-grained, as has been done for the activity of large neural populations, to ask which effective variables and couplings survive the flow~\cite{meshulam2019}. But coarse-graining acquires physical meaning only once the degrees of freedom being coarse-grained have been identified. Cross-order Laplacian schemes extend Laplacian renormalization~\cite{villegas2023} to diffusion on simplices and define scale transformations for higher-order structure~\cite{nurisso2025}. In most of the models considered here, however, the dynamical variables live on nodes, while simplices specify the interactions among them. Unless the simplices carry relevant slow degrees of freedom, coarse-graining them settles nothing: the question is not whether higher-order structure changes across scales, but whether interaction order survives as a relevant variable of the effective theory.

In short, before a group interaction is credited with producing new phenomena, three questions should be answered: \emph{What changes? What survives at long scales? What can be inferred from data?} These questions carry a condition: a comparison is informative only if the two descriptions use comparable physical resources. Replacing a group interaction at the price of a vastly enlarged auxiliary state space establishes representational equivalence, not explanatory equivalence.

An argument that ended in negation would be a poor one. There are places where higher-order structures can deliver naturally what a pairwise language becomes strained to express. Seen this way, the cases discussed above already point to them. One is when it makes the organizing structure explicit. In the Newman--Moore model, the three-spin product and the lattice geometry jointly generate fractal subsystem symmetries and glassy dynamics without quenched disorder~\cite{newman1999}. Likewise, the higher-order Laplacian used by Lucarini \emph{et al.} reveals relaxation scales that locate a Griffiths-like regime~\cite{lucarini2026}. Higher-order structure can also simply be the better mechanistic language: a formal reduction buys little if it costs memory, hidden variables and a proliferating auxiliary state space~\cite{neuhauser2020,llabres2026}. Social systems provide a classical example of this last kind: from Simmel's triads onward, a third actor changes the roles available to the other two, even where the resulting dependence can be represented as a nonlinear response to pairwise inputs~\cite{simmel1908,granovetter1978}.

The useful question is therefore not whether higher-order interactions are real---they plainly are---but what they buy that a lower-order description does not: what survives because of them, what becomes visible through them, and what becomes simpler when they are kept.

\section{Epilog}
We suspect the deeper issue is methodological. The renormalization group did not become the central tool of statistical physics by adding degrees of freedom; it became central by removing them. Its content is the claim that progress consists in identifying which variables survive coarse-graining and which do not.  A field that responds to every new dataset by inventing a new microscopic rule is doing the opposite: it accumulates. The result is not a theory but a bestiary, and a bestiary has no theoretical reach---each entry explains only itself.

Anderson's essay is usually read as a defense of emergence against reductionism~\cite{anderson1972}. It is worth recalling the other half of his argument, which is easier to forget: emergent laws are interesting because they are \emph{simpler} than the microscopic ones and largely independent of them. Universality is what turns ``more is different'' from a slogan into a scientific statement. It is also what makes a physicist's model worth more than a fit. The same essay closes with the apocryphal exchange between Fitzgerald and Hemingway~\cite{anderson1972}, and it makes the point better than any argument. Fitzgerald: ``The rich are different from us.'' Hemingway: ``Yes, they have more money.'' Higher-order systems may yet prove different from pairwise ones. So far, what has been shown is that they have more legs. Telling the two apart is the research program for the near future.

\section{acknowledgments}
We thank our colleagues in the higher-order community, whose work we criticize here only in the sense in which physicists ordinarily criticize work they take seriously.

\bibliographystyle{apsrev4-2}
\bibliography{refs}

\end{document}